\documentclass[]{elsart}

 \usepackage[english]{babel}
 \usepackage[T1]{fontenc}
 
\usepackage{graphicx}
 
\usepackage{epsfig}
\usepackage{wrapfig}
 
\usepackage{amssymb,amsmath,amsfonts}
 
\usepackage{url}

\usepackage[numbers,square,compress,sort]{natbib}

\begin{document}

\begin{frontmatter}
 
 

\title{Generating of an electric potential on the Moon by Cosmic rays and Solar Wind?}

%

\author{M.J. Simon and J.~Ulbricht\corauthref{cor}}
\address{ETH Z\"urich, Institute for Particle Physics, CH-8093 Z\"urich,
Switzerland}
\corauth[cor]{Corresponding author}
\ead{masimon@phys.ethz.ch}
\ead{ulbricht@phys.ethz.ch}

\begin{abstract}
 
We investigate the possibility that the Moon develops an electric
potential originating from the impinging particles on the Moon
from cosmic rays and solar wind. The investigation includes
all experimental data of the flux of charged particle for energies
higher than 865 eV available from
Apollo missions, satellites and balloon experiments in publications
or from the Internet in 2008.
A fictive electric potential of the Moon was calculated if the Moon
material is an isolator for the Moon solar side and lee side,
if the Moon material is a conductor for the whole Moon surface, and
if the Moon is located in the geomagnetic tail of the Earth.
The calculation for these four cases results in positive electric
potentials of the Moon of 1789~V, 261~MV,  1789~V, and 96~MV.
This is originated from the unequal distribution of positive and
negative charges in the plasma of the cosmic rays and solar wind
impinging on the Moon. As the cosmic rays arrive from deep space, these
findings would imply a charge imbalance in the cosmos. This is in
distinct conflict with a charge neutral universe. We suggest 
searching for a so far not measured low energy negative flux of
charged particles in the cosmos or an interaction between charged
objects in the universe with the vacuum.
 
\end{abstract}
 
 
\end{frontmatter}
 
\section{Introduction}
\label{sec:Introduction}

That our universe is baryon asymmetric as a whole is established by
observations in contemporary cosmology. If large regions of matter and
antimatter coexist, annihilation would occur at the borders between them.
This annihilation would generate a diffuse $\gamma$-ray background
disturbing the cosmic microwave radiation and light element abundance.
No such effect was observed so far~\cite{Chechetkin82,Khlopov87,Khlopov99}. Analysis
of this problem~\cite{Rujula} for a baryon symmetric universe demonstrates
that the size of regions exceed 1000 Mpc, being comparable with the
modern cosmological horizon. Even under the circumstances that still
some small anti-matter regions are possible~\cite{Sasha} and the
anti-Helium search of AMS-I~\cite{HeAMS1} does not exclude
anti-Helium for the whole universe, the probability that the universe
is baryon symmetric is low.
 
A baryon symmetric universe would be charge symmetric because
matter and anti-matter are distinguished prevailing 
by charge and anti-charge.
As a consequence the sum over all charges in such an universe must be
zero. But also our baryon asymmetric universe is controlled by charge
conservation. S. Orito et.al. discuss possible temporal
charge non conservation in
cosmological models based on theories with a higher dimensional space leading
to a charged universe~\cite{CanUnivBeCharged}.
A cosmology of a charged universe and the total
electric charge and mass of an elliptical universe is investigated in
ref.~\cite{CosmologyChargedUniverse,Asanov}.
Consequences of a charged universe for primordial Helium production
and massive vector fields are evaluated in
ref.~\cite{PrimHelium,MassVectorfield}.
All these investigations conclude that the
charge imbalance in our universe is not significant.
In fact also our
baryon asymmetric universe is with high probability globally not charged.

The baryons in our universe are with high abundance concentrated in galaxies
which are composed of objects like stars and planets and clouds of gas and dust. If these objects
would carry an excess of positive or negative charges, electric fields
between these objects must exist.

In contemporary cosmology the existence of cosmological magnetic
fields are well established~\cite{MacroscopicBfield1,MacroscopicBfield2}.
Since large-scale electric fields have not been observed,
and since one would not expect cosmological electric currents
and charge distributions, cosmological electric fields
are assumed to be zero~\cite{MacroscopicEfield1,MacroscopicEfield2}.
Theoretical considerations of the enormous linear dimensions
and the electrical conductivity of the completely ionized gas 
inside the galaxies support the assumption that 
interstellar electric fields are minute~\cite{IntergalacticEfield1,IntergalacticEfield2}.
In proposals for a Bussard ramjet interstellar electric fields
of $ 1.6 \cdot 10^{-19} $~V/m~\cite{IntergalacticEfield3,IntergalacticEfield4}
are in discussion. The interplanetary electric field is caused by ions leaving the Sun.
The ions initially flow along and parallel to the magnetic field
of the Sun, further outwards the azimuthal component of the
magnetic field becomes more influential. Protons get deflected
to the south and electrons to the north. This results in an electric
field that compensates the magnetic forces~\cite{InterplanetaryElectricField1}. 
Substantial theoretical and experimental
investigations on interplanetary electric fields are performed.
Studied are models of stationary electromagnetic fields in
interplanetary space with finite conductivity 
in ref.~\cite{InterplanetaryElectricField2} (With more references therein).
Interplanetary electric fields under disturbed conditions
are investigated in ref.~\cite{InterplanetaryElectricField3}.
Magnetic storms are discussed 
e.g. in~\cite{InterplanetaryElectricField4,InterplanetaryElectricField5}. WIND observations
of electrostatic fluctuations found a total potential
drop between the Sun and the Earth to be about 300 V to
1000~V~\cite{InterplanetaryElectricField6,InterplanetaryElectricField7}.
Polar cap ionospheric electric fields tends to saturate
at approximately 45 - 50~mV/m~\cite{InterplanetaryElectricField8}.
In summary no substantial cosmic, intergalactic or interplanetary
electric fields are observed.
This supports the hypothesis that all objects in the cosmos
are electrical neutral. Which requires that the electric
interaction between these
objects can only be achieved by a neutral plasma.
This electric interaction is performed
by cosmic rays, intergalactic medium, interstellar medium
and the solar wind.

The cosmic rays are energetic particles originating from outer
space. They are composed of approximately 90\% protons, 9\% helium and
1\% electrons~\cite{DefinitionCosmicRays}.
 
The intergalactic
space between the galaxies is nearly free of dust and debris, very close
to a total vacuum. A minor amount of baryons is stored in this
intergalactic medium, a rarefied plasma of equal numbers of electrons 
and protons~\cite{intergalacticplasmaI,intergalacticplasmaII}.
The majority of the content of
the intergalactic medium is electrical neutral but some
fraction in particular the H~II regions are ionized.
The interstellar medium what is located between the stars within a
galaxy consists of an extremely dilute mixture of ions, atoms, molecules,
large dust grains, cosmic rays, and galactic magnetic
fields~\cite{componentsinterstarmedium}. 
 
If we move further on to the solar system, the interstellar wind
has to pass the bow shock, heliopause and the solar wind termination
shock. These boundaries together with the magnetosphere of the
planets are shielding for example the Earth substantially
from low energy interstellar wind. This is true in particular in
the direction of the movement of our solar system
with respect to the center of the Milky Way. 
This shielding is not effective at higher energies,
for example the Alpha Magnetic Spectrometer AMS
measured a lepton spectrum of the kinetic energy higher than  0.2~GeV
at altitudes near 380 km above the Earth surface~\cite{AMSleptons}.
 
The dominating part of particles inside the solar system
is the solar wind. A stream of charged particles (plasma)
ejected from the upper atmosphere of the Sun. It consists
mostly of protons and electrons with energies of approximately
1~keV. The solar wind creates the Heliosphere, a bubble in the
interstellar medium surrounding the solar system. The investigation
of the solar wind has a long tradition. It reaches from Richard C.
Carrington in 1859 suggesting the concept of streams of particles
flowing outward from the Sun, to the first direct observation
by the Soviet satellite
Luna~1~\cite{Solarwind1,Solarwind2,Solarwind3,Solarwind4,Solarwind5,Solarwind6}
and the solar
plasma experiment of Neugebauer et al. using the Mariner~2 
spacecraft~\cite{Mariner2}.
 
The aim of our paper is to confront the hypothesis, that
all objects in the cosmos are electrical neutral, with
the extensive experimental data, observations, and studies of the
energy distribution and
particle flux of the cosmic rays, intergalactic medium, interstellar
medium, and the solar wind collected in the last years. If all these fluxes
impinge on an object in space and if they are a neutral plasma,
the sum over the time of all these positive and negative charged
fluxes must add up to a total charge zero. The Earth Moon
has no atmosphere or intrinsic magnetic field, and consequently
its surface is bombarded almost all time with the full set of
particles. For this reason we use the Moon as ideal example to
test the hypothesis of an electrical neutral object. 
An excess of positive or negative
charge flux would establish after some time an electric
potential (electric field) on the Moon.
We do not in consider photoemission from the Moon surface
because photoelectrons affect the electric potential of the Moon only 
locally in the near vicinity of the Moon~\cite{Photoemission}.

The article is organized as follows. We first select four test scenarios
to investigate, is the flux of cosmic rays and solar wind behaving
like a neutral plasma. We calculate a fictive electric
potential of the Moon originated from these fluxes.
Accordingly to these test scenarios we calculate
four electric potentials of the Moon on its orbit circling
the Earth and conclude.

\section{Definition of four scenarios to test the neutral plasma hypothesis}
\label{sec:Moon Orbits}
The Moon orbits the Earth in approximately 27.3 days, its
spin is synchronized with the orbit cycle. As a consequence
the same half of its surface points permanently to the
Earth. On its path the Moon is exposed to cosmic rays at all positions.
The low energy particles flux from 
intergalactic and interstellar media is suppressed by the
bow shock, heliopause and termination shock outside the
solar system. The most intense particle stream bombarding
the Moon is the solar wind.
 
On its orbit  half of the Moon surface is fully exposed
to the solar wind plus the cosmic rays. The other half
is illuminated only by the cosmic rays. 
The solar wind is shadowed if the Moon
passes through the geomagnetic tail of the Earth magnetic field.
In this location only cosmic rays impinge on the Moon.
These various
exposure conditions of the Moon surface open the possibility
to test four neutral plasma scenarios. First, if the material
of the Moon behaves like a nonconductor, the charged stream of the
solar wind plus the cosmic rays on the solar side of the Moon
must integrate over the time to a total charge zero. Second, the
charged stream of cosmic rays on the solar wind lee side
must fulfill the same condition. Third, in the more likely case that the Moon
material behaves like a conductor, the solar wind plus cosmic rays
on the solar  side of the Moon plus the cosmic rays on the
solar wind  lee hemisphere must behave like a neutral plasma. Fourth, if the
Moon passes the geomagnetic tail, almost the full surface of the Moon is
exposed to the cosmic rays only and the neutral plasma condition
should be still valid.
\subsection{Lunar surface solar wind observations of Apollo 12 and Apollo 15}
\label{sec:Lunar surface solar wind observations}
\begin{figure}[htb]
\centerline{\psfig{file=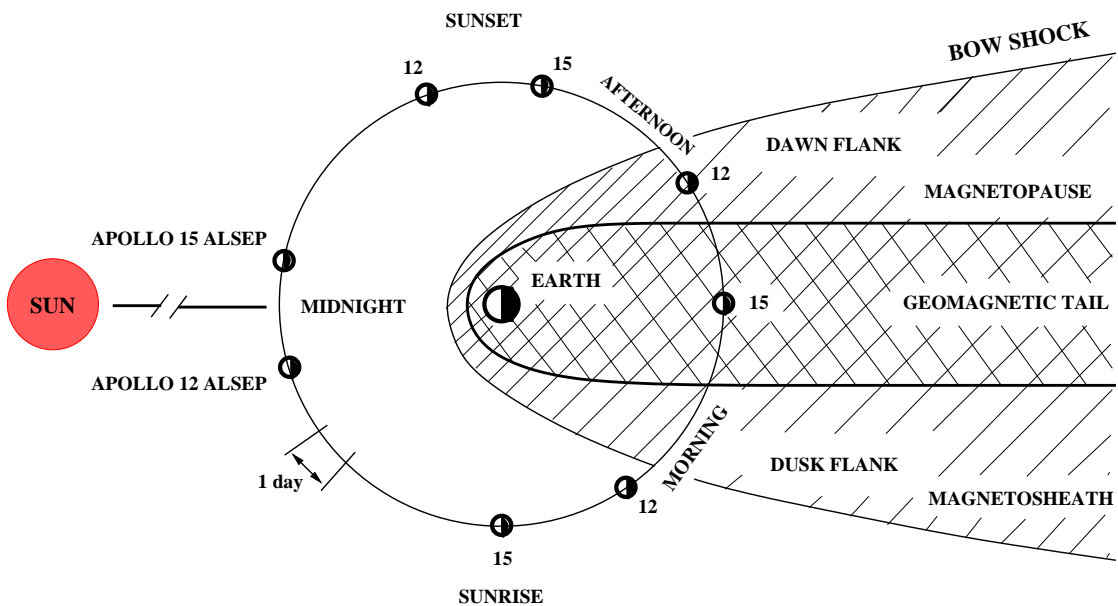,width=16cm}}
\caption{\it Approximate location of the Moon for noon, sunset, midnight,
and sunrise at the Apollo 12 and Apollo 15 sites. Picture taken from 
Ref.~\cite{Apollo12and15} Fig.~3 }
\label{moonposition}
\end{figure}

The lunar surface solar wind observations at the Apollo 12
and Apollo 15 sites~\cite{Apollo12and15} open the possibility 
to test these four neutral plasma conditions. The
approximate location of the Moon for
sunrise, sunset, morning, noon, afternoon and night
for an observer on the Moon
at the Apollo 12 and Apollo 15 sites
is shown in Fig.~\ref{moonposition}.

The lunar latitudes and longitudes of the Apollo 12 and 15 sites
are $3\,^{\circ}{\rm S}$, $23\,^{\circ}{\rm W}$ and
$26\,^{\circ}{\rm N}$, $4\,^{\circ}{\rm E}$, respectively.
As a consequence, the sunrise and sunset of Apollo 12 and
Apollo 15 is at different path locations. If Apollo 15 is
inside the geomagnetic tail, Apollo 12 is in the dawn flank
of the magnetopause. For completeness also shown is
an example of the Apollo 12 an Apollo 15 ALSEP 
(Apollo Lunar Surface Experiments Package) location~\cite{ApolloALSEP}.
\subsection{Collected flux data of Apollo 12, Apollo 15, balloons and satellites experiments}
\label{sec:Collected flux data}
If $\psi$ is the lunar longitude of the solar
direction, then Apollo~15 took proton data between $ \psi \sim -100\,^{\circ} $ in the
morning until $ \psi \sim 90\,^{\circ} $ in the afternoon. No data are taken
when the Moon is inside the geomagnetic tail or during lunar night because
the ion fluxes are below the solar wind spectrometer
threshold~\cite{Apollo12and15}.
The proton flux is approximately stable between $ -70\,^{\circ} < \psi <
-23\,^{\circ}$ and $ 18\,^{\circ} < \psi < 70\,^{\circ}$. We assume
this is true for the hemisphere of the Moon pointing to the Sun.
We imply this data in the total data set, as proton
flux for the test cases one and three. Out of the given
density of 8~protons/cm$^3$ and the bulk velocity of 405~km/s, which is
equivalent to 856~eV, the spectral flux was calculated to be
$6.02\cdot10^8$~(m$^{-2}$~s$^{-1}$~sr$^{-1}$~eV$^{-1}$).
%

We did not use the electron flux at the lunar surface measured from the Apollo
experiments~\cite{ApolloElectronen} because our interest is focused on the
total
electron flux impinging on the Moon from the solar wind. On the Moon surface
local electrical and magnetic fields together with photo electrons overlay
the detection of the pure solar wind electrons~\cite{LUNARPhotoElectrons}.
Collected are data of charged cosmic rays measured at the top of the 
atmosphere
of the Earth (TOA, about 100~km) or higher or at the L1 libration point.
Fluxes
measured at the TOA or higher or fluxes which are at least corrected to TOA
are collected
because these fluxes are not altered by the Earth atmosphere.
The L1 libration point is a point of Earth-Sun gravitational equilibrium
about $1.5\cdot10^6$~km from Earth and $148.5\cdot10^6$~km ($\sim$ 1AU) from
the Sun. It is assumed that the same fluxes measured at TOA or at the L1
libration impinge on Moon since Earth, Moon and L1 libration point are in
near vicinity. Fluxes of neutrons and neutral atoms are not collected; they
are not of interest since these particles carry no charge. Further, fluxes
of particles trapped in the Van Allen Radiation Belt (VARB) are excluded.
The VARB is a torus of charged particles encircle the Earth, held in place by
the Earth magnetic field. Since this is a local phenomenon of the Earth and
since the Moon orbit does not cross the VARB, the particles in the VARB
may not be counted to the flux of the particles which impinge on the Moon.  
 
Data are taken from following experiments:
AMS~\cite{AMS2}: The Alpha Magnetic Spectrometer (AMS) measured the
proton spectrum in the kinetic energy range (0.2 - 200)~GeV during the space
shuttle flight STS-91 in June 1998. The space shuttle flew at an altitude of
about 380 km so that the data are free from atmospheric corrections.
BESS 2000~\cite{Asaoka}: The Balloon-borne Experiment with a
Superconducting Spectrometer (BESS) measured among others the antiproton
spectrum in the kinetic energy range (0.18 - 4.2)~GeV during several balloon
flights in 1999 - 2000 (and earlier). For simplicity only the data from 2000
are taken. The experiments were carried out in northern Canada at altitudes
above 34 km and the data are corrected to TOA.
CAPRICE94~\cite{Boezio} and CAPRICE98~\cite{Boezio1}: The Cosmic
Antiparticle Ring Imagine Cherenkov Experiment (CAPRICE) was a balloon-borne
experiment which collected data of the cosmic ray electron and positron
spectra with kinetic energies about (0.46 - 43.6)~GeV in August 1994. Data
of cosmic ray proton and helium spectra with (15 - 150)~GeV/n were collected
in May 1998. Both times the altitude of the balloon was about 37~km and the
data are corrected to TOA.
CRN~\cite{Muller}: The Cosmic Ray Nuclei Detector (CRN) measured the cosmic
ray
nuclei from carbon to iron with kinetic energies per nucleon beyond 1~TeV/n.
The measurement was done during the Space-lab-2 mission on the Space Shuttle
Challenger in 1985 which flew at an altitude of about
320~km~\cite{Wikipedia1}. The data are therefore free from atmospheric
corrections.
HEAO-3-C2~\cite{Engelmann}: The French-Danish cosmic ray experiment C2 was
launched in 1979 on board the NASA High Energy Astrophysics Observatory-3
(HEAO-3) satellite. It measured the fluxes of nuclei from beryllium to
nickel in the energy range from 0.6~GeV/n to 35~GeV/n at an altitude of
about 500 km~\cite{NASA2} so that the data are free from atmospheric
corrections. Measurements expired in June 1980 after a failure in the high
voltage system.
IMAX~\cite{Menn}: The Isotope Matter-Antimatter Experiment (IMAX) measured
the cosmic ray proton and helium spectra in the energy range (0.2 -
200)~GeV/n in July 1992. It was a balloon-borne experiment carried out in
northern Canada at an altitude of about 36~km, the collected data are
corrected to TOA. Orth et al.~\cite{Orth}:
Cosmic ray nuclei from lithium to iron in the
energy range (2 - 150)~GeV/n were measured by a balloon-borne
superconducting
magnetic spectrometer. The measurement was done in September 1972 at an
altitude of about 35.5~km. The data are corrected to TOA. Most of the data
of this experiment are adequately covered by the HEO-3-C2 and CRN
experiments, so that only the flux of lithium, which is not measured by
HEO-3-C2 or CRN, is used for the further analysis.
{RUNJOB}~\cite{RUNJOB}: The RUssia-Nippon JOint Balloon-program (RUNJOB)
measured cosmic ray protons with (10 - 500)~TeV, helium nuclei with  
(3 - 70)~TeV/n and iron nuclei with (1 - 5)~TeV/n. The data are results of
measurements from 1995 to 1996, where the balloon flew at an average
altitude of about 32~km over the trans-Siberian continent.
TRACER~\cite{TRACER}: The Transition Radiation Array for Cosmic Energetic
Radiation (TRACER) is a balloon experiment which measured heavier cosmic ray
nuclei (O, Ne, Mg, Si, S, Ar, Ca and Fe) at high energies. The balloon had a
successful flight in Antarctica in 2003 and reached an average altitude of
about 38~km. The data are corrected to TOA.

Further data can be found on several Internet web pages which provide data
for download. Under the address \url{http://www.srl.caltech.edu/ACE/} data
from the instruments of the ACE (Advanced Composition Explorer) satellite
can be downloaded. The ACE satellite orbits the L1 libration point since
1998. It has several instruments on board, from which the following were
used:
The Cosmic Ray Isotope Spectrometer (CRIS), which measures nuclei from boron
to nickel of cosmic rays with energies about (50 - 450)~MeV/n. The
Solar Isotope Spectrometer (SIS). SIS measures nuclei from helium to nickel
over the energy range (10 - 100)~MeV/n. ULEIS, the Ultra Low Energy Isotope
Spectrometer, measures ion fluxes from helium through nickel from about
20~keV/n 
to 10~MeV/n. The Electron, Proton and Alpha Monitor (EPAM), which
measures ions from (0.05 - 5)~MeV/n and $e^-$ with (0.04 - 0.31)~MeV. Only
electron data are used from EPAM since the ion measurements do not clearly
separate the particle species.
The flux data can be downloaded in different time intervals, mostly are
available: hourly, daily, and Bartels Rotation averages. The latter are averages
over 27 days, approximately the period for one solar rotation. This time
average was chosen for the download of all data from the beginning of the
measurement until the 70 DOY (day of year) 2008 for ULEIS and SIS and until
the 97 DOY 2008 for CRIS and EPAM.
\newline
Data of the IMP-8, SOHO and WIND satellites can be downloaded from
\url{http://cdaweb.gsfc.nasa.gov/cdaweb/}.
IMP-8 (Interplanetary Monitoring Platform) had an elliptical orbit around
the Earth with apogee and perigee distances of about 45 and 25 earth radii
so that it does not cross the VARB. The spacecraft was in the solar wind for
7 to 8 days of its every 12.5 day orbit. In October 2001 the mission of IMP
8 was terminated after 29 years of successful measurements~\cite{NASA:IMP8}.
The proton and helium flux data of this experiment are used; they can be
downloaded for the whole measurement period in 30 minutes averages.
The Solar and Heliospheric Observatory mission (SOHO) orbits the L1
libration point since 1996. In June 1998 contact to SOHO was lost for about
130 days, but since October 1998 SOHO is still active until
today~\cite{NASA:SOHO}. The electron flux data of this experiment are used.
Although SOHO collects data until today, the data can be downloaded only for
the
period from December 1995 until January 2002 in 5 minutes averages. 
WIND was started in November 1994. It was placed in a double-lunar swing by
orbit near the ecliptic plane, with apogee from 80 to 250 Earth radii and
perigee of between 5 and 10 Earth radii for the first nine months of
operation. After nine months WIND was steered in a small orbit around the
L1 liberation point where it remains until today~\cite{NASA:WIND}. The
electron flux data of WIND since 1996 until the end of 2006 are used. The
data from November 1994 until the end of 1995 are not used because they
contain probably particles of the VARB due to the spacecraft orbit. The
electron data of the WI\_K0\_3DP instrument can be downloaded in time
averages of about 90 seconds and the data of the WI\_ELSP\_3DP instrument
are available in time averages of about 50 seconds.
\subsection{Data processing} 
\label{sec:data processing}
Most experiments specify explicitly in which energy interval and at which
average energy the fluxes are measured. If the average energy is given, this
value is used for the further analysis. If only the energy interval is
given, the average energy is calculated along the recommendation of the
EPAM, SIS and ULEIS experiments: The root of the product of lower and upper
limit of the energy intervals is taken as average energy.

Particle fluxes in space vary in short and long time scales, mainly due to the
solar activity. Especially fluxes with energies below $\sim\!0.1$~GeV/n,
which are measured by ACE, IMP-8, SOHO and WIND,
change the intensity substantially as a function of time. They contain
partially solar wind particles and solar energetic particles as low
energetic galactic and anomalous cosmic rays~\cite{mewaldt:165}. They exhibit
approximately the expected eleven-year-cycle of the Sun. 
Our interest was focused on the average charged particle flux
impinge on the Moon.
We ignored for this reason all these short term and long term variations
of the low energetic fluxes and took the mean value over a longer time scale.
 
The total flux of one particle species is obtained by integration of the
spectral flux of this particle  over its energy range. To perform the
integration, power functions in the form of
\begin{equation}%
f_{n}(E_k) = a \cdot 10^{-9} (E_k 10^{-9}/\textrm{1~eV})^{-b}, \ \ [a]=
\frac{1}{\textrm{m$^2$ s sr eV}}
\label{eq:fitkct}%
\end{equation}%
were calculated between two data points. $E_k$ is the kinetic
energy of the particles inserted in units of eV.
The fit-parameter $b$ is dimensionless and  
$a$ has the unit of (m$^{-2}$s~$^{-1}$~sr$^{-1}$~eV$^{-1}$).
Multiplication with $10^{-9}$ is used for
numerical reasons. The index $n$ stands for a particular
particle species.
The measurement of the particle flux as function of the energy
of one particle species contains usually more than two data points.
We introduced for this reason the global fit function $F_n(E_k)$.
The function
$F_n(E_k)$ is generated by connecting the various single fit functions
$f_{n}(E_k)$.
Outside of the data range, where no data points are available, the global
fit functions $F_n(E_k)$ are set to zero.
Tests with alternative fit functions between more than two data points
as discussed in ref.~\cite{Gaisser,mewaldt:227,Petrov}
deliver very similar results after integration.  We use 
therefore for simplicity the two data point approach.
 
\begin{figure}[htb]%
\begin{center}
\centerline{\psfig{file=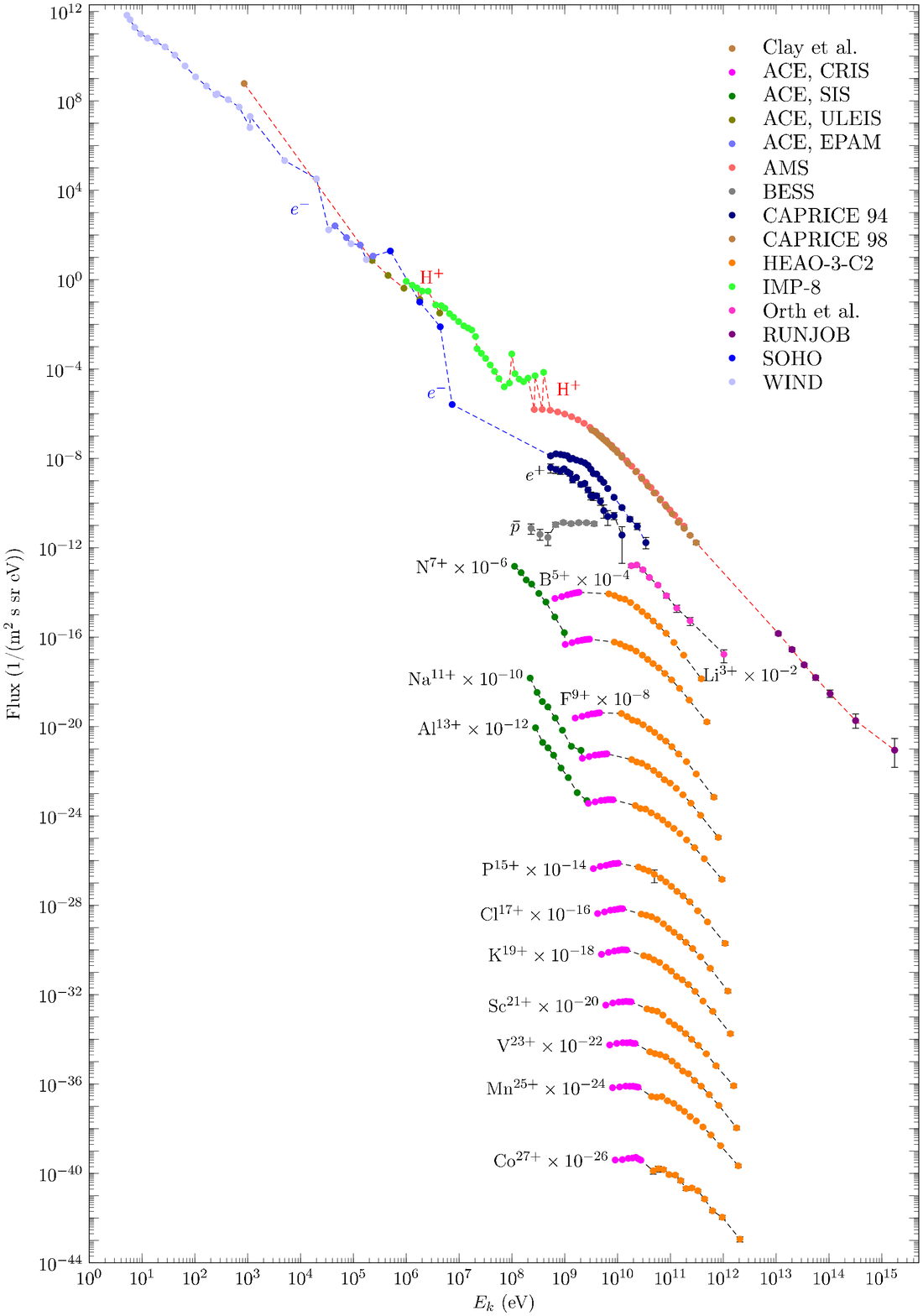,width=12.5cm}}
\end{center}%
\caption{Fluxes of odd charged particles at the top of the atmosphere or
above. To facilitate the visibility of the single fluxes in the
graphic the fluxes are shifted by factors of $ 10^{-n} $.
The fit
functions are indicated with black (for electrons with blue, for protons
with red) dashed lines.}%
%
\label{fig:fluxes}
\end{figure}

Fluxes of all odd charged particles are shown in Fig.~\ref{fig:fluxes}. A
similar picture would be obtained if even charged particles are plotted. For
simplicity all ions are treated as fully ionized, which is true for
high energetic particles.
The spectral fluxes of heavier nuclei than protons are small and
have only minor influence on the total (spectral) charge flux.
Inspection of Fig.~\ref{fig:fluxes} exhibits the fact that
in the whole measured energy range of the protons 
flux, the positive charged flux is larger than the negative 
charged flux.

\section{Calculation of an electric potential of the Moon}
\label{sec:Calculation of the electric potentials}
The impinging particles from the cosmic rays and solar wind
which are able to reach the surface of the Moon will
be accumulated in the Moon material.
If the positive and negative charged particle flux as function of
energy and time is equal, the Moon surface will integrate
the flux spectrum to a total charge zero. The present available
data of the particle fluxes do not allow to investigate the whole
energy range $0 < E_{k} < \infty$ because electron flux data below
5.2~eV and proton flux data below 856~eV are not available.
Inspecting Fig.~\ref{fig:fluxes} a smooth extension of the
particle data flux to $E_{k} \sim 0$ point in the direction
of an excess of positive charged particles not only at high 
energies but also at energies close to zero. So far, this
assumption is not confirmed from measurements which
limits our investigation of an excess of positive charged
particle flux to energies higher than  856~eV.
The excess of positive charges
will result in a positive electric field and potential in the
environment of the Moon. The numerical value of this potential
will be a function of the kinetic energy and the flux of the impinging charged
particles. The leading parameter of the kinetic energy 
is the relative velocity between the Moon surface and the
particles. The kinetic energy of the particles get transformed
in the electric field of the Moon to a potential energy
if the particle moves against the repelling force of the field.
With increasing  positive potential
an increasing repelling force between Moon and further impinging positive charged
particles will develop. 
Without relative velocity no repelling electric field could be generated.
Depending
on the potential of the Moon, low energetic particles will not be able
any more to reach the Moon surface, the
low energy part of the particle flux in Fig.~\ref{fig:fluxes}
will be cut away. On the other hand,
the whole energy spectrum from
$ E=0 $ to $ E= \infty $ of the negative charged particle flux will reach
the Moon. With increasing potential the
Moon will in particular for very low energetic negative particles act 
like an attractor to collect all negative
charges in its environment. The positive electric potential of the Moon
will increase and after a certain
time the flux of positive charges and negative charges reaching the 
Moon will be equal and a time independent stable positive electric
potential will be established.

The particle density $ n $ of the cosmic rays and the solar wind is
in the environment of the Moon 
between $ n = 3 $ cm$^{-3} $  and $ n = 8 $ cm$^{-3} 
$~\cite{PlasmaDensityMoon1,Apollo12and15,PlasmaDensityMoon2} 
depending of the solar activity.
The mean free path of electron-ion, ion-ion
and electron-electron
scattering is in the highly diluted plasma for
energies of $ 1 $~eV  or $ 1 $~keV  and $ n=8 $~cm$^{-3} $
approximately $ 1.6 \cdot 10^{3} $ respectively
$ 1.6 \cdot 10^{7} $ Moon radii~\cite{F-F-Chen}.
The Debye length is for
$ 1 $~eV  or $ 1 $~keV  and $ n=8 $~cm$^{-3} $ between
$ 2.6 $~m respectively $ 83 $~m~\cite{F-F-Chen}.
A low energy charged particle ($ 1 $~eV  to $ 1 $~keV)
moving in direction of the Moon surface
will not undergo any scattering after it passed a distance to the
Moon of $ 1.6 \cdot 10^{3} $  respectively  $ 1.6 \cdot 10^{7} $
Moon radii. The interaction of the particle is restricted
only to a possible electric field and the negligible magnetic field
of $ 16.6 $ nT~\cite{PlasmaDensityMoon1}. We use for this reason
a single particle solution to calculate the path of the charged
particles to the Moon surface. We take into account the interaction
of an electric field (potential) of the Moon on
the path of the charged particles.

If the particles are very close to the Moon surface of approximately
seven lunar radii, in the lunar wake an electric negative potentials
of about 400 V was measured~\cite{Ogilvie}. A generation of an
electric field in this close environment originating from photoemission
is discussed in ref.~\cite{Photoemission}.
It would be possible
to generate double layers, a region of non-neutral plasma
consisting of two adjacent space charge layers, one
positive and one negative charged. Such layers would 
affect in particular low energy (eV) cosmic rays and
solar wind on its path to Moon surface. These particles
would also change the conditions for the double layers
because the particle flux from the cosmic ray and 
solar wind seems not to be charge neutral.
This photoemission do not
contribute to the total charge integral of the particles
impinging on the Moon in a distance of $ 1.6 \cdot 10^{3} $
Moon radii or further outside. Also the mean free path of the
cosmic ray and solar wind is much larger as seven Moon radii.
We ignore for this reason these local electric fields for our
calculations of the path of the cosmic rays
and solar wind to the Moon surface because our interest is
focused on the total integrated charge in a large
distance from the Moon.
\subsection{Low energy cut from measured data range}
\label{sec:Energy cuts}
After our extensive study of the data of charged 
particle fluxes in Sec.~\ref{sec:Collected flux data}, 
it is evident from Fig.~\ref{fig:fluxes} that data for low energy
charged particles are not available. Measurement of e.g. the electron flux
between $ 0 < E_k < 5.2 $~eV and of the proton flux
between $ 0 < E_k < 856 $~eV are not present.  

We use for this reason a low energy cut for all data at
\begin{equation}
 E_{k}^{cut} = 856~\textrm{eV}.
\label{eq:LowEnergyCut}
\end{equation}
This cut limits our calculation to energies 
$ E_k $ between
\begin{equation}
 856~\textrm{eV} \leq E_k < \infty.
\label{eq:EnergyWindow}
\end{equation}
\subsection{Attractive and repulsive electric 
            forces on charged particles induced 
            from an electric potential of the Moon}
\label{sec:Change}
If the Moon is exposed to an excess of a positive
charge flux $ F(t,E,\sigma)$ in the energy range of
Eq.~\eqref{eq:EnergyWindow}, after some time an excess
of positive charges would be collected.
The amount of this charge $Q_{C}$ is
a function of the collection time $t$,
the energy of the impinging particles $E$,
and the surface $\sigma$ of the Moon which is exposed to the flux.
Including $ d\sigma=R^{2}~d\Omega $, $R$ the Moon radius
and d$\Omega$ as the solid
angle,  $Q_{C}$ is the integral of Eq.~\eqref{IntegratedCharge}.
\begin{equation}
  Q_{C} =
\int{\int{\int{F(t,E,\sigma)~\textrm{d}t}~\textrm{d}E}~\textrm{d}\sigma}
\label{IntegratedCharge}
\end{equation}
The charge $Q_{C}$ will develop a potential
$\phi_{C}$ of the Moon.
If we assume the Moon is a perfect sphere,
the size of this potential on the Moon surface is
a function of the charge $Q_{C}$
of the excess flux in Eq.~\eqref{IntegratedCharge} and
the radius $ R $ of the Moon.
\begin{equation}
  \phi_{C} = \frac{1}{4 \pi \varepsilon_{0} R } Q_{C}
\label{IntegratedPoteialC}
\end{equation}
We introduce a single particle solution
to calculate the evolution of a time dependent
potential $\phi_{C}$ of the Moon under the 
influence of the charged particle flux ~\cite{Diplomarbeit}.
For the calculation it is taken into account that
same-sign charged particles are repelled  and different-sign charged
particles are attracted if the Moon has a potential $\phi_C > 0$.
Positive charged particles will be deflected
by $ \phi_{C} $, and depending on their energy and
impact parameter they will not reach the surface of the Moon.
Contrary, negative charged particles will be attracted by
$ \phi_{C} $ even for large impact parameter and 
large energies.

\begin{figure}[htb]
\vspace{5.0 mm}
\begin{center}
\centerline{\psfig{file=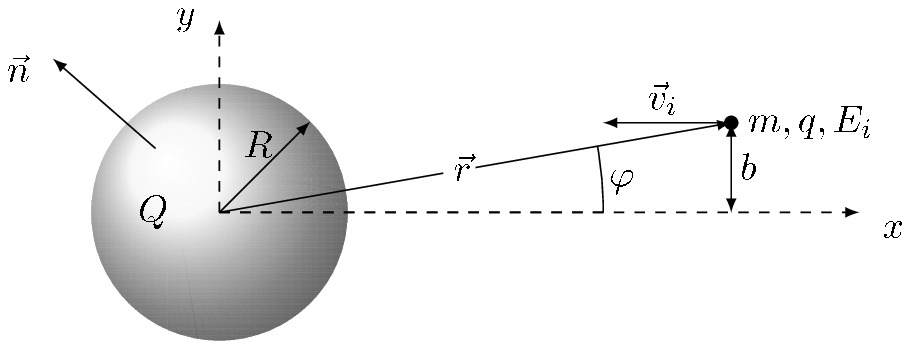,width=10cm}}
\end{center}
\caption{A particle with rest mass $m$, charge $q$, velocity $v_i$ and
initial energy $E_i$, moves from infinity towards to the Moon
of radius $R$. 
The position vector of the particle is $\vec{r}$, $b$ denotes the impact
parameter, $n$ an unit vector on the sphere and $\varphi$ the angle between
the $x$-axis and the position vector $\vec{r}$}%
\label{fig:sphere}%
\end{figure}

We begin with a time independent
potential $\phi$, which is necessary to repel same-sign charged and to attract
different-sign charged particles. $\phi$ can be calculated easily if Moon is treated as
perfect sphere. The calculation is performed in a semi-relativistic approach
because bremsstrahlung,
which is emitted from relativistic charged particles 
if they are decelerated or accelerated from the potential
of the Moon, is ignored.
In Fig.~\ref{fig:sphere} is shown how a particle with rest mass $m$, charge
$q$, velocity $v_i$ and initial energy $E_i$,  moves
from  infinity towards a sphere.

If energy conservation and conservation of angular momentum $L$ is taken
into
account, $ \phi $ can be calculated as in 
Eq.~\eqref{eq:deflectionPotential}
\begin{equation}
\phi = \frac{E_i}{q}\left(1- \frac{1}{E_i R}\sqrt{L^2 c^2 + R^2 m^2 c^4}
\right).
\label{eq:deflectionPotential}
\end{equation}
The (initial) kinetic energy $ E_{k,i} $ of a particle is calculated 
as in Eq.~\eqref{eq:kinEnergy}
\begin{equation}
E_{k,i} = E_i - m c^2.
\label{eq:kinEnergy}
\end{equation}
The angular momentum $ L $ can be expressed as in
Eq.~\eqref{eq:angularmomentum} under the approximation that 
the initial velocity $v_i$ of the particle stays constant.
%
\begin{equation}
L = m \gamma v_i r \sin \varphi = m \gamma v_i b, \ \ \  \gamma =
\frac{1}{\sqrt{1-(v_i/c)^2}},
\label{eq:angularmomentum}
\end{equation}
By inserting Eq.~\eqref{eq:angularmomentum} in
Eq.~\eqref{eq:deflectionPotential}, it is possible to
define an attraction and repulsion function $g(E_{k,i},\phi)$ in
Eq.~\eqref{eq:ImpactParameter}.
\begin{equation}
g(E_{k,i},\phi)\doteq \frac{b}{R} = \sqrt{\frac{(E_{k,i} - q \phi)^2 + 2 m
c^2(E_{k,i} - q \phi)}{E_{k,i}(E_{k,i} + 2 m c^2)}}
\label{eq:ImpactParameter}
\end{equation}
In Eq.~\eqref{eq:ImpactParameter} $q\phi$ is the electrostatic potential energy; 
in the case of attraction $q\phi< 0$ and for 
repulsion $q\phi> 0$.

\begin{figure}[htb]
\vspace{10.0 mm}
\begin{center}
\rotatebox{270}{
\epsfig{file=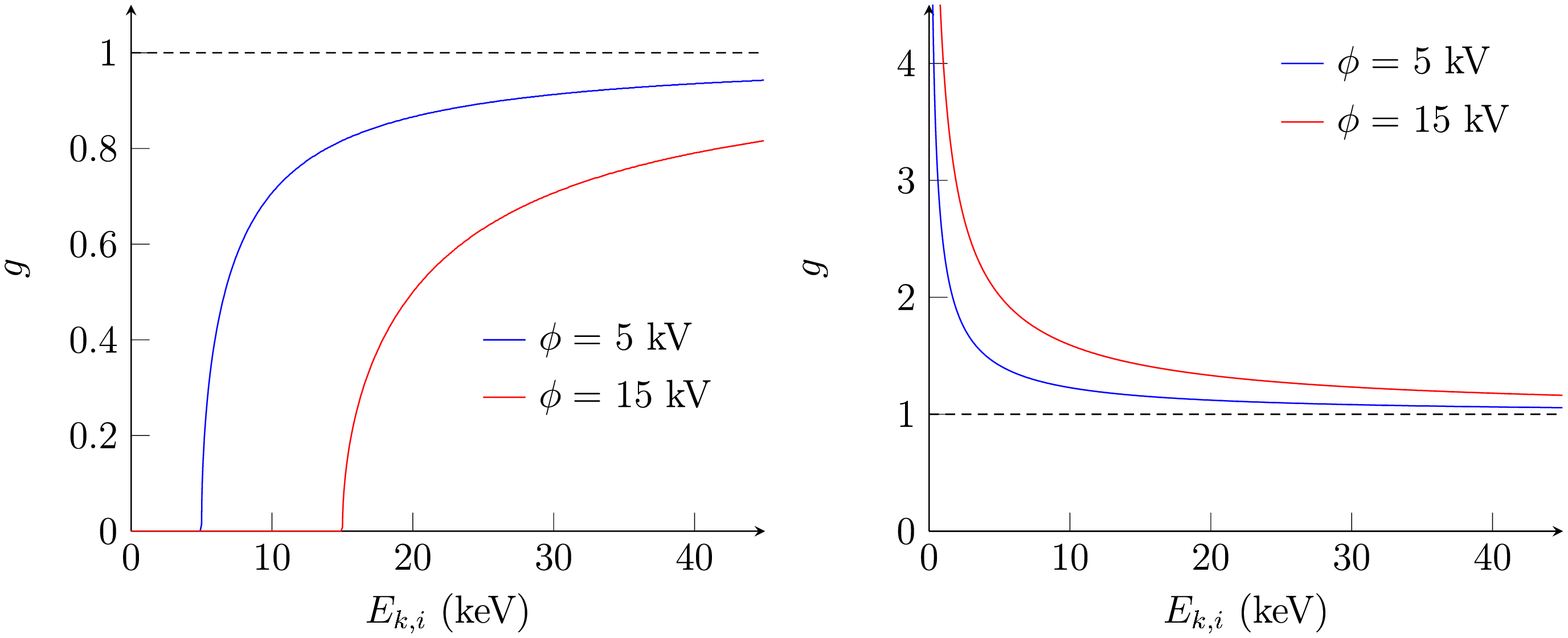,width=5.0cm}}
\end{center}
\vspace{5.0 mm}
\caption{Repulsion (left side) and  attraction (right side) function
$g(E_{k,i},\phi)$ as a function of $E_{k,i}$ for a
potential of $\phi$ = 5 kV and $\phi$ = 15 kV.
On the left side for the repulsion case only particles
between $ 0 \le~ g(E_{k,i},\phi)~<~ 1 $ and
$ E_{k,i}~> $ 5 keV or 15 keV
are able to approach
the surface of the Moon.
For the attraction case (right side) particles between
$ \infty~>~ g(E_{k,i},\phi) \ge 1 $ will impinge on
the surface of the Moon. Not displayed are particles
between $ 0 \le  g(E_{k,i},\phi) \le 1 $ which reach in
addition the surface of the Moon.
Totally in the attraction case all
particles which fulfill the condition
$ \infty ~>~g(E_{k,i},\phi) \ge 0 $ will impinge on
the surface of the Moon. }
\label{fig:ImpactParameter}%
\end{figure}

In the case the forces between Moon and charged particle are repulsive $(q\phi> 0)$,
 the function
$ g(E_{k,i},\phi)=b/R=0 $ for $ 0 \le E_{k,i} \le q\phi $; 
for $ E_{k,i} > q\phi $
is $ 0 < g(E_{k,i},\phi) < 1 $
and for  $E_{k,i} \rightarrow \infty $
is $ g(E_{k,i},\phi)=1 $.
The physical range for collection of
charged particles on the Moon of the function
$ g(E_{k,i},\phi) $ is displayed in
Eq.~\eqref{eq:EnergyRangRepulsPlusFuntion}.
\begin{equation}
  0~\le~g(E_{k,i},\phi) = \frac{b}{R}~<~1
\label{eq:EnergyRangRepulsPlusFuntion}
\end{equation}
An example for the  function $ g(E_{k,i},\phi) $
as function of the energy $ E_{k,i} $
for a potential of the Moon of
$\phi$ = 5 kV and $\phi$~= 15 kV is
displayed in Fig.~\ref{fig:ImpactParameter} left side.
Particles with $E_{k,i}<q\phi$ are not able to reach the
surface of the Moon. Particles with $E_{k,i}>q\phi$
impinge the surface of the Moon if their impact
parameter fulfills the condition
$ b \le g(E_{k,i},\phi) \cdot R $.

In the attraction case $(q\phi< 0)$, 
two groups of particles which are able to
impinge on the surface of the Moon have to be considered. The first group
belongs to particles with an impact parameter $ \infty > b > 1 $ and
the second to an impact parameter $ 1 > b \ge 0 $.

In the first case,
for $E_{k,i}=0$ is
$ g(E_{k,i},\phi)=\frac{b}{R} \rightarrow \infty $, for $E_{k,i} > 0$ is
$ g(E_{k,i},\phi) > 1 $ and for  $E_{k,i} \rightarrow \infty $
is $ g(E_{k,i},\phi) \rightarrow 1 $. This part of the physical
range for collection of
charged particles on the Moon of the function
$ g(E_{k,i},\phi) $ is displayed in
Eq.~\eqref{eq:EnergyRangRepulsMinusFuntion}.
\begin{equation}
  \infty > g(E_{k,i},\phi) = \frac{b}{R} \geq 1
\label{eq:EnergyRangRepulsMinusFuntion}
\end{equation}
An example for the 
function $ g(E_{k,i},\phi) $
for a potential of the Moon of
$\phi$~= 5 kV and $\phi$~= 15 kV as a function
of the kinetic energy for the first group of particles is
displayed in Fig.~\ref{fig:ImpactParameter}, right  side.
All particles in the energy region of
$ 0 \leq E_{k,i} < \infty $ and the physical
range of $ g(E_{k,i},\phi) $ according
Eq.~\eqref{eq:EnergyRangRepulsMinusFuntion} 
are impinging on the surface of the Moon.
The second group of particles in the environment of the Moon
which are located in the range of the impact parameter
$ 0 \leq b \leq R $ will reach the surface of the Moon
in the attractive case 
for all energies between
$ 0 \leq E_{k,i} < \infty $.
We include in the following calculations
all particles in the range
of the impact parameter $ 0 \leq b < \infty $.
\subsection{Electric potential of the Moon}
\label{sec:Electric potential of the moon}
To investigate the time dependence of an
electric potential of the Moon $\phi_C(t)$,
we use the semi-relativistic approach of 
Eq.~\eqref{eq:deflectionPotential} and set
\begin{equation}
\phi = \phi_{C}.
\label{eq:PhicPhi}
\end{equation}
This ensures that all same-sign
charged particles with an energy $E_{k,i} < q\phi_C$ can not reach the Moon surface
anymore and that same-sign charged particles with an energy $E_{k,i} > q\phi_C$ can
reach only restricted parts of the Moon surface, 
as soon as the Moon potential $\phi_C$, which is caused 
by the collected charge $Q_C$, is
not equal to 0. This ensures also that different-sign charged 
particles between $0 < E_{k,i} < \infty$ with $b > R$ are attracted by the
Moon potential

We introduce in Eq.~\eqref{IntegratedPoteialC} a time dependence 
shown in Eq.~\eqref{eq:PotentialMoon}
\begin{equation}
\phi_C(t) = \frac{1}{4 \pi \epsilon_0 R} ( Q_C(t) + Q(t)_{unid} )
\label{eq:PotentialMoon}
\end{equation}
$ Q_C(t) $ is the time dependent charge collection of the Moon.
For the so far unmeasured charged particle flux below 856~eV of
Eq.~\eqref{eq:LowEnergyCut} and a mechanism besides our single particle
solution to collect charges we introduce $ Q(t)_{unid} $. 
We set $ Q(t)_{unid}=0 $ and will discuss this term later.
To stress the time dependence of $ Q_C(t) $ we rewrite
Eq.~\eqref{IntegratedCharge} to Eq.~\eqref{eq:ChargeCosmicRay}.
\begin{equation}
 Q_C(t) = \int I_C ( t ) \textrm{d}t.
\label{eq:ChargeCosmicRay}
\end{equation}
According to Eq.~\eqref{eq:fitkct} and Eq.~\eqref{IntegratedCharge},
$I_C(t)$ is the induced charge flux of the total
cosmic ray flux $ F_n(E_{k,i}) $ impinging on the Moon.
For times $t \sim 0$ the attraction of different-sign 
charged particles and the repulsion of same-sign charged particles can be
neglected. The current can then be calculated as in Eq.~\eqref{eq:Current}:
\begin{equation}
 I_C (t \sim 0 ) = \sum_{n} \left( \Delta \Omega
        \int \limits_{0}^{\infty}
         R^2 \cdot q_n \cdot F_{n} (E_{k,i}) \textrm{d}E_{k,i} \right)
\label{eq:Current}
\end{equation}
The index $n$ of the sum runs over all measured contributions 
of the charged particle fluxes
$ F_{n} (E_{k,i}) $ of Eq.~\eqref{eq:fitkct} we discussed in
chapter~\ref{sec:Collected flux data} and displayed in
Fig.~\ref{fig:fluxes}. The charge of the particles is  $q_n$ and
$\Delta \Omega $ is the solid angle under which the Moon is hit by
charged particles.
Inserting Eq.~\eqref{eq:Current} and Eq.~\eqref{eq:ChargeCosmicRay}
in Eq.~\eqref{eq:PotentialMoon},
leads to the Moon potential 
for $ t \sim 0 $:
\begin{equation}
\phi_C(t \sim 0) = \sum_{n} \left(  \frac{ \Delta \Omega }{4 \pi \varepsilon_0 }
          \int \limits_0^{t \sim 0} \int \limits_{0}^{\infty}
          R \cdot q_n \cdot F_{n}(E_{k,i}) \textrm{d}E_{k,i}  \textrm{d}t'
\right).
\label{eq:PotenialMoonWithoutgfunc}
\end{equation}
For $ t > 0 $ the Moon will develop
an electric field and repulsive and attractive forces between Moon and
charged particles will develop. 
Following our investigation of the
potential of the Moon $ \phi_C(t) $ as a function of
the impact parameter $ b $ in chapter~\ref{sec:Change},
we include this interaction in
our calculation by replacing
in Eq.~\eqref{eq:PotenialMoonWithoutgfunc} $ R $ by the
impact parameter $b=R\cdot g(E_{k,i},\phi_C) $ of
Eq.~\eqref{eq:ImpactParameter}. The value of the impact parameter $ b $
as a function of the Moon potential $ \phi_C(t) $ is a leading
parameter of the value of the charged particles flux impinging on
the Moon surface.
The final result for the Moon potential is then given by:
\begin{equation}
\phi_C(t) = \sum_{n} \left(  \frac{ R \Delta \Omega  }{4 \pi \varepsilon_0 }
          \int \limits_0^t \int \limits_{0}^{\infty}
          g(E_{k,i},\phi_C(t')) \cdot q_n \cdot F_{n}(E_{k,i})
\textrm{d}E_{k,i}  \textrm{d}t'  \right)
\label{eq:PotenialMoon}
\end{equation}
It is possible to solve Eq.~\eqref{eq:PotenialMoon} approximatively
with a recursion formula if the derivative of this expression is calculated
and expressed as difference quotient. Neglecting the limit of the
difference quotient, which is valid for small time differences $\Delta t =
t_m - t_{m-1}$, and replacing $\phi_C(t)$ with $\phi_C(t_{m-1})$  in the
attraction and repulsion function $g(E_{k,i},\phi_C(t))$, which is valid
since 
$\phi_C(t_{m-1}) \approx \phi_C(t_m)$  is an approximation for
$\phi_C(t)$, leads then to the calculable recursion formula
Eq.~\eqref{eq:PotenialMoonIteration}:
\begin{equation}
\begin{split}
\phi_C(t_m) & = \sum_{n} \left(
\frac{R \Delta \Omega }{4 \pi \varepsilon_0}\Delta t \int
\limits_{0}^{\infty} g(E_{k,i},\phi_C(t_{m-1})) \cdot q_n \cdot
F_{n}(E_{k,i}) \textrm{d}E_{k,i} \right)+ \\ 
& \ \ \ + \phi_C(t_{m-1}) 
\end{split}
\label{eq:PotenialMoonIteration}
\end{equation}
For the iteration we chose for the initial value
$t_0 = 0$ and  $\phi_C(t_0)=0$.
Choosing  $\phi_C(t_0)\neq 0$ would
describe an increase or decrease of the
potential of an initially charged Moon if
the impinging particle flux changes its intensity.
In the following section we use Eq.~\eqref{eq:PotenialMoonIteration}
to study the four test scenarios discussed in
Sec.~\ref{sec:Moon Orbits}.
\section{Four test scenarios of electric potentials 
         of the Moon on its orbit to circuit the Earth}
\label{sec:ResultDecission}
To test neutral plasma conditions of the flux of charged particles 
in space it would be in the first approach adequate to inspect
Fig.~\ref{fig:fluxes}. Obviously at higher energies the positive
flux of charged particles exceeds the negative charged flux. At low
energies in the  eV  regime we have only data from the electrons.
The first data point for the positive flux at 856~eV is still
above the electron flux. Considering the uncertainty at low energies,
a qualitative tendency of a bigger positive particle flux 
compared to the negative flux is more likely.
In the following sections we discuss
the in Sec.~\ref{sec:Moon Orbits} introduced
four scenarios.

\subsection{Test~1 of neutral plasma conditions for an insulating hemisphere of the Moon facing the Sun}
\label{sec:TEST1}
\begin{figure}[htb]
\vspace{-45.0 mm}
\begin{center}%
\rotatebox{270}{
\centerline{\psfig{file=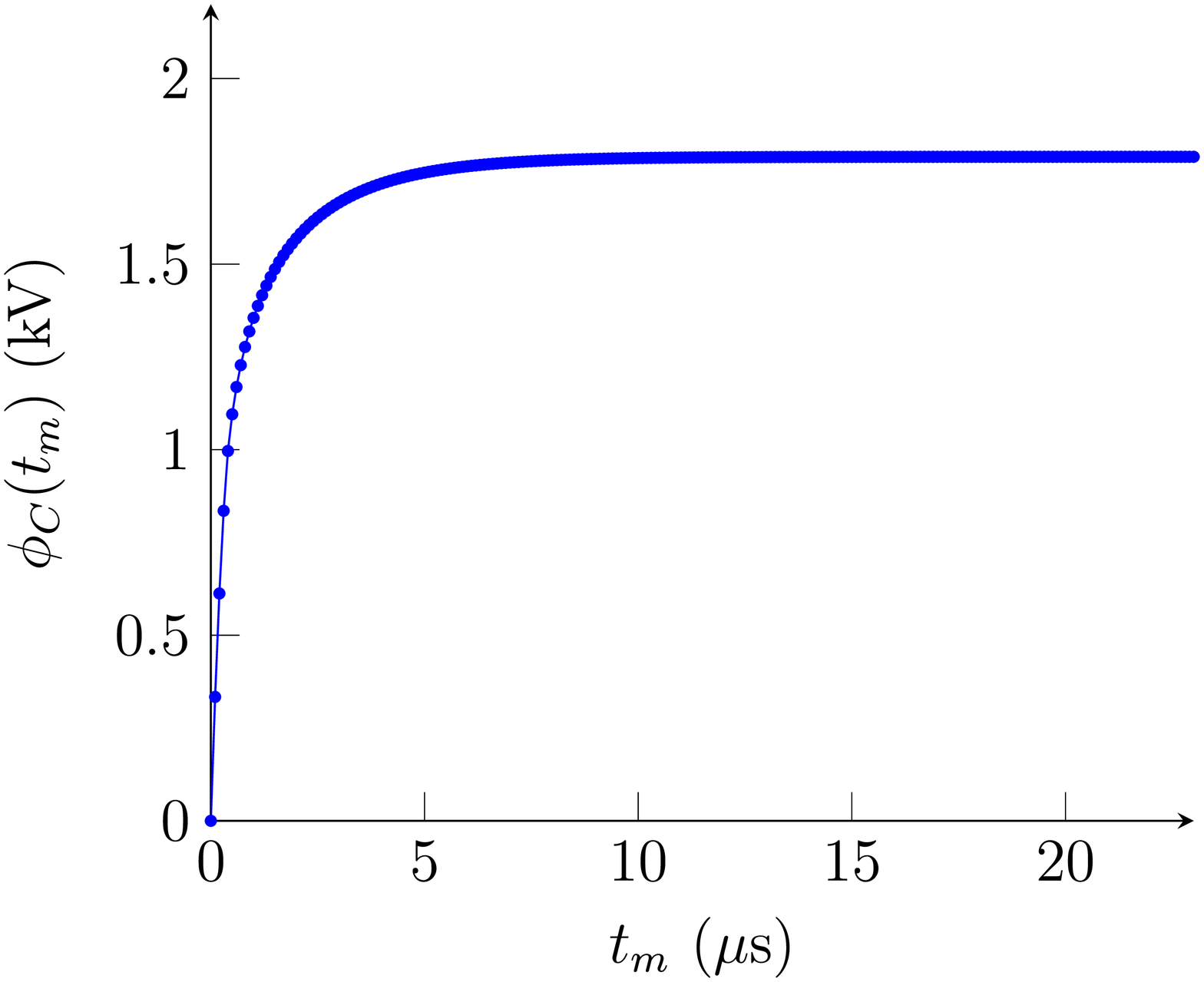,width=3.8cm}}}
\end{center}%
\vspace{-44.0 mm}
\caption{ Test~1. Increase of the electric potential of an isolator Moon
whose solar side
is exposed to solar wind plus cosmic rays.}
\label{fig:MoonPotentialIncreaseTest1}
\end{figure}

We use Eq.~\eqref{eq:PotenialMoonIteration} to calculate
the time evolution of a potential on the Moon solar side. The
calculations are performed including the constraints, that
the charged current of the
solar wind plus the cosmic rays impinge only the Moon solar side 
and that no charge exchange will take place between solar and solar lee side
of the Moon because we assume that
the material (rocks) of the Moon behave like an isolator.
The time dependence of the potential $\phi_C(t_m)$
is displayed in Fig.~\ref{fig:MoonPotentialIncreaseTest1}.

The Moon would develop in approximately 13.4~$\mu$s a positive potential of
1789~V. After this time the 
repulsion effect for positive 
and attraction effect  for negative charged particles, discussed in Sec.~\ref{sec:Energy cuts}, would stabilize
the potential $\phi_C$ at this level. As a consequence, the hemisphere of the
Moon pointing to the Sun would be
on this potential as long as the Moon is outside of the Earth geomagnetic tail.

For the calculation the
initial value $\phi_C(t_0)$ and 
the initial time $t_0$ were chosen to be zero, $\Delta \Omega$ was set 
to $2 \pi$ which takes into account that we use
only the half surface of the Moon.
To generate a smooth function of $ \phi_C(t_m) $ in
Fig.~\ref{fig:MoonPotentialIncreaseTest1},
$\Delta t$ was set to 0.1~$\mu$s.
Independently of all test scenarios, we assume that the flux of the
solar wind impinging the surface of the Moon is directed to the Sun. We set
for this reason for the solar wind the solid angle to $2\pi$.
%
%
\subsection{Test~2 of neutral plasma conditions for an 
insulating hemisphere of the Moon in the lee of the solar wind}

\label{sec:TEST2}
Eq.~\eqref{eq:PotenialMoonIteration} is used to calculate
the time evolution of a potential on the Moon solar lee side. The
constraints are the same as in test~1 with the exception, that only
cosmic rays impinge this side.
 
\begin{figure}[htb]
\vspace{-40.0 mm}
\begin{center}%
\rotatebox{270}{
\centerline{\psfig{file=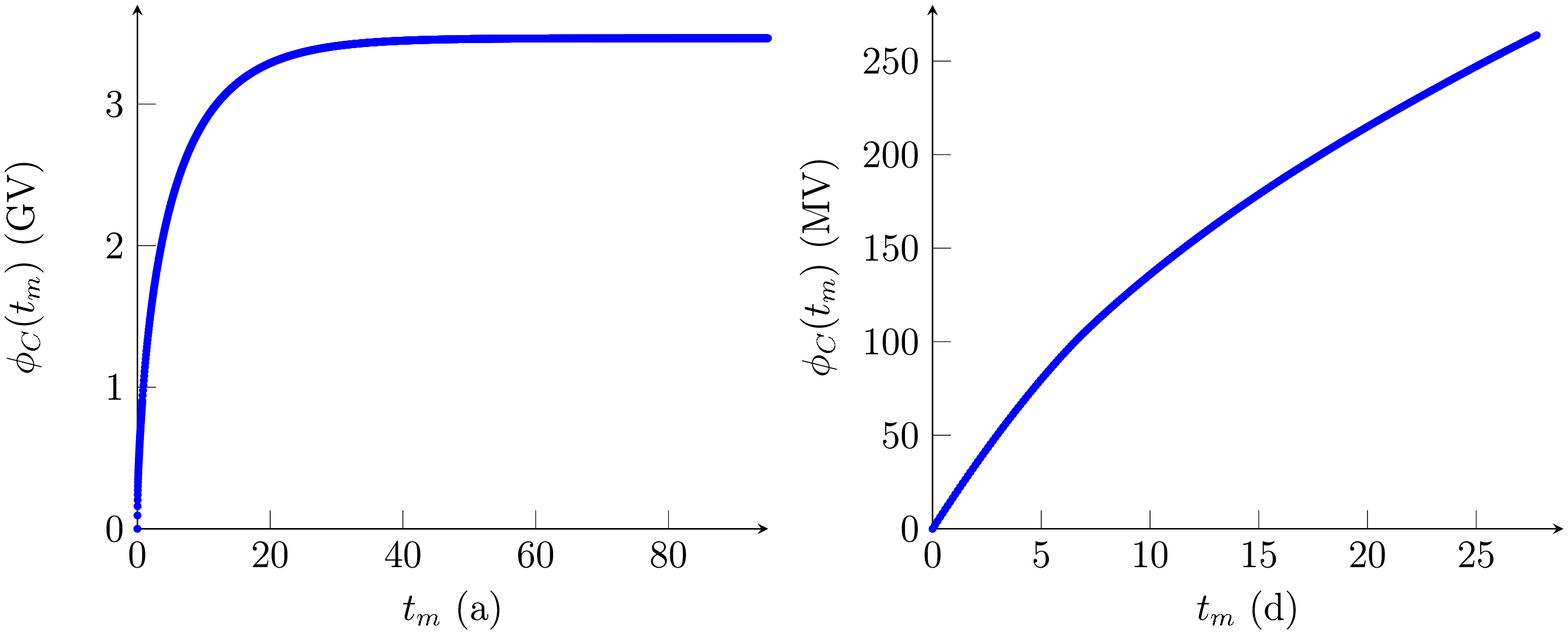,width=5.0cm}}}
\end{center}%
\vspace{-40.0 mm}
\caption{Test~2. Increase of the electric potential of an isolator Moon if
its solar lee side
is exposed only to cosmic rays (left side for a period of about 100
years;
 right side for a period of about 27.3 days).}
\label{fig:MoonPotentialIncreaseTest2}
\end{figure}

The time dependence of the potential $ \phi_C(t_m) $
calculated with Eq.~\eqref{eq:PotenialMoonIteration} is displayed in
Fig.~\ref{fig:MoonPotentialIncreaseTest2}

The Moon would develop in approximately 78 years a positive potential of
3.465~GV. After this time the potential $\phi_C$ would be stable
at this level. Due to the Moon spin, the conditions which part of the
surface of the Moon is exposed to the two possible particle 
fluxes change per day. For this reason it is only possible to
calculate an upper limit for the potential if it is assumed that
one half of the surface of the Moon is exposed e.g. 27.3 days to cosmic
rays. In this time the Moon would reach a potential of approximately
261 MV as shown on the right sid
of Fig.~\ref{fig:MoonPotentialIncreaseTest2}.

For the calculation 
of Eq.~\eqref{eq:PotenialMoonIteration}
the start value $\phi_C(t_0)$ and 
the start time $t_0$ were chosen to be zero and $\Delta t$
was set to $4.5 \cdot 10^5$~s.
For the calculation of the iteration shown on 
the right side of Fig.~\ref{fig:MoonPotentialIncreaseTest2},
$\Delta t$ was set to $10^4$~s, for which 
the influence of the attraction and repulsion effect is seen.
$\Delta \Omega$ was set in all iterations 
to $2 \pi$ which takes into account that the particles impinge  
only on one half of the Moon surface. 
The separation of cosmic rays from solar wind was performed
with an energy cut at 0.1~GeV/n.
All global fit
functions $F_n(E_{k,i})$ are set to zero for energies below
0.1~GeV/n. This ensures that almost only cosmic rays are taken
into account for the calculation.

Compared to test~1
substantial differences in the electric potential would occur
at the surface of the Moon at the transition between solar and 
and solar lee side.

\subsection{Test~3 of neutral plasma conditions for
a conductor Moon which is hit by solar wind and cosmic rays}

\label{sec:TEST3}
In the more likely case that the Moon
material behaves like a conductor~\cite{lunarElectricalConductivity1,
lunarElectricalConductivity2,lunarElectricalConductivity3,
lunarElectricalConductivity4},
the flux of the solar wind
plus cosmic rays on the Moon solar side plus the cosmic rays on the Moon
solar lee side are added together.

\begin{figure}[htb]
\vspace{-40.0 mm}
\begin{center}%
\rotatebox{270}{
\centerline{\psfig{file=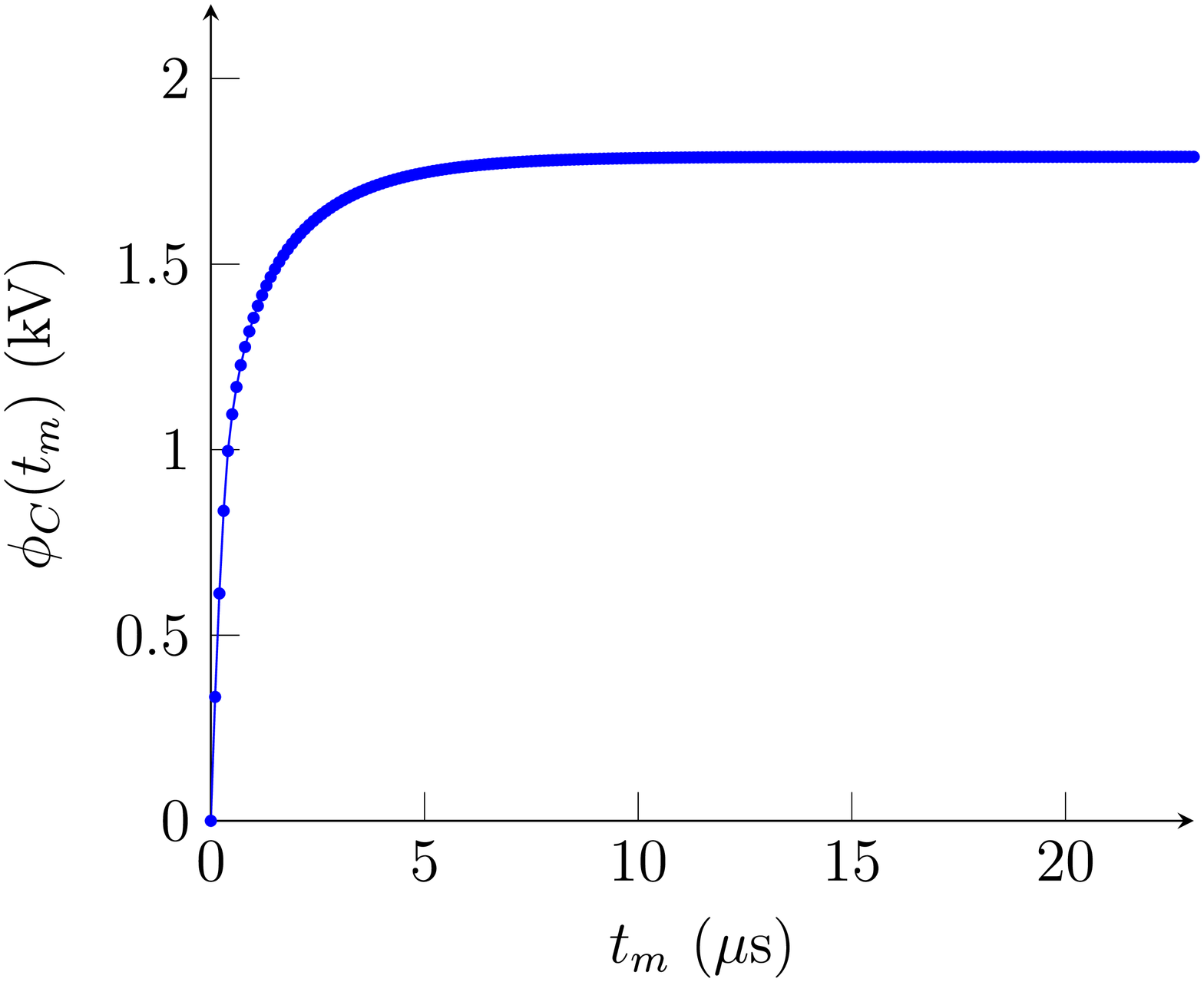,width=4cm}}}
\end{center}%
\vspace{-40.0 mm}
\caption{Test~3. Increase of the electric potential of an conductor Moon
whose solar side is exposed to solar wind plus cosmic rays
and whose solar lee side is exposed only to cosmic rays.}
\label{fig:MoonPotentialIncreaseTest3}
\end{figure}

The time dependence of the potential $ \phi_C(t_m) $
calculated with Eq.~\eqref{eq:PotenialMoonIteration} is displayed in
Fig.~\ref{fig:MoonPotentialIncreaseTest3}

The Moon would develop in approximately 13.4~$\mu$s a positive
potential of 1789~V. After this time the potential $\phi_C$ 
would stabilize on this level as long the Moon is outside of the
geomagnetic tail of the Earth.

In the calculation we used for the Moon solar side
the unchanged global fit functions to add together the
flux of the solar wind and cosmic rays. On the Moon solar lee
side we used the altered global fit functions, which are set 
to zero for energies below 0.1~GeV/n to include only
cosmic rays.
$\Delta \Omega$ is set
to $2 \pi$ which takes into account that solar wind plus cosmic
rays on the solar side of the Moon
and the cosmic rays on the
solar lee side of the Moon impinge in each case on one half
of Moon surface.
$\phi_C(t_0)$ and
$t_0$ were chosen to be zero and $\Delta t$ to 0.1~$\mu$s.

The result demonstrates the absolute dominance of the solar wind
compared to the cosmic rays.

%
%
\subsection{Test~4 of neutral plasma conditions 
            for a conducting  Moon in the geomagnetic tail of the Earth}
\label{sec:TEST4}
\begin{figure}[htb]
\vspace{-40.0 mm}
\begin{center}%
\rotatebox{270}{
\centerline{\psfig{file=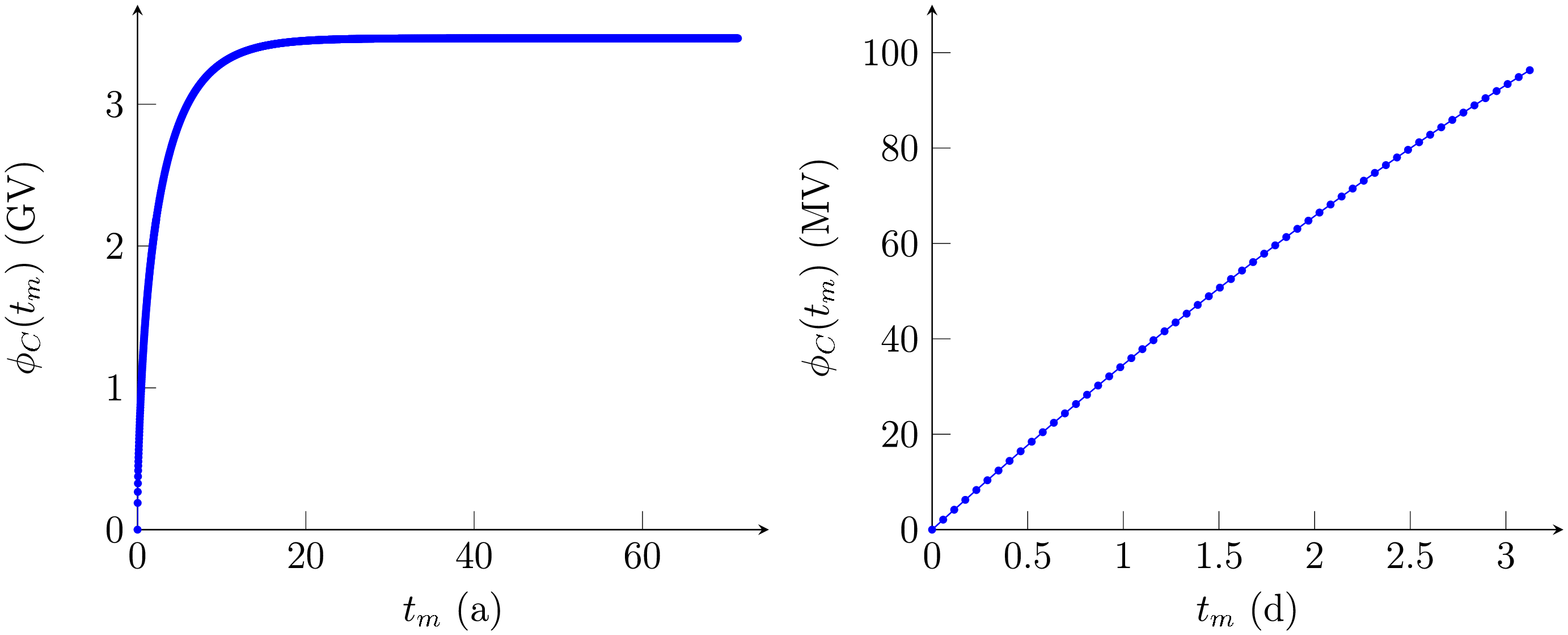,width=5cm}}}
\end{center}%
\vspace{-40.0 mm}
\caption{Test~4. Increase of the electric potential of a conductor
Moon whose full surface is exposed only to cosmic rays
(left side for a period of about 70~years; right side for a period of about
3.1~days when the Moon is in Earth geomagnetic tail).}
\label{fig:MoonPotentialIncreaseTest4}
\end{figure}

If the Moon passes the geomagnetic tail of the Earth, the full surface of the
Moon is exposed to cosmic rays only.
The time dependence of the potential $ \phi_C(t_m) $
calculated with Eq.~\eqref{eq:PotenialMoonIteration} is displayed in
Fig.~\ref{fig:MoonPotentialIncreaseTest4} left side.

The Moon would develop in approximately 39 years a positive potential of
3.465~GV, which will be stable after this time.
As the Moon is only approximately 3.1 days of its orbit in the
geomagnetic tail of the Earth, the potential would approach
96~MV as shown on the right side
of Fig.~\ref{fig:MoonPotentialIncreaseTest4}.

To ensure that almost only cosmic rays
are taken into account we set in the  calculation
all global fit
functions $F_n(E_{k,i})$ to zero for energies below
0.1~GeV/n. 
We use as initial values $\phi_C(t_0)=0$ and
$t_0=0$. The time $\Delta t$ we set to $4.5 \cdot 10^5$~s
for the calculation of the iteration shown on
the left side of Fig.~\ref{fig:MoonPotentialIncreaseTest4} and
$\Delta t$ was set to 5000~s for the right side of
Fig.~\ref{fig:MoonPotentialIncreaseTest4}.
To respect the isotropy of the cosmic rays~\cite{PTUSKIN}
we chose the solid angle of $4\pi$.
\subsection{Time dependence of the
            potential of the Moon orbiting the Earth}
\label{sec:Comparisonof tests}
\begin{figure}[htb]%
\vspace{-30.0 mm}%
\begin{center}%
\rotatebox{270}{
\centerline{\psfig{file=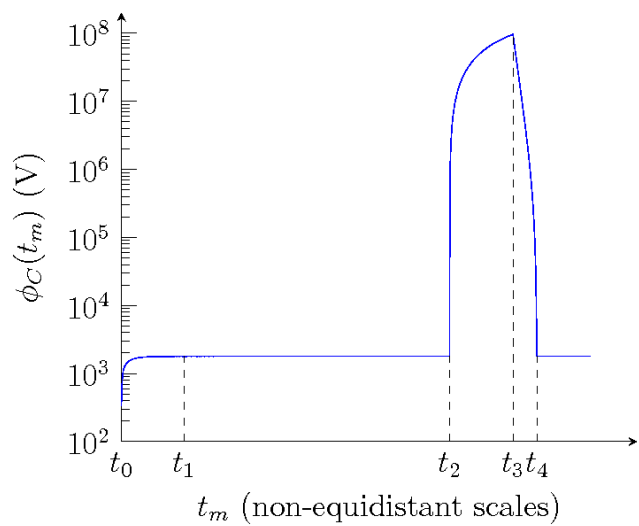,width=7cm}}}
\end{center}%
\vspace{-30.0 mm}%
\caption{Development of Moons potential for a conducting Moon on its
27.3 day orbit.}%
\label{fig:MoonOrbitMonth}%
\end{figure}

In Fig.~\ref{fig:MoonOrbitMonth} we combined the time dependence
of the potential of the Moon on its 27.3~day
orbit circling the Earth. We used the more realistic case
of a conductor Moon from test 3 and test 4.

To visualize the rapid changes of the potentials,
non-equidistant scales were chosen on the time axis.
According test~3, between $t_0$ and $t_1$
the initially uncharged Moon would be charged to a
potential of 1789~V in 13.4~$\mu$s. Its
solar side is exposed to solar wind plus cosmic rays
and its solar lee side to cosmic rays.
The Moon would stay at this potential until it enters after
approximately 24.2 days the Earth geomagnetic tail at $t_2$.
The potential would increase to about 96~MV
in the next 3.1 days. At $t_3$ the Moon
leaves the geomagnetic tail of the Earth and is
exposed again to the conditions of test 3. Under the 
influence of the high fluxes of the solar wind the
potential would decrease rapidly in the next 1~ms until it
reaches at $t_4$ the value of 1789~V again.
We used for the potential decrease between $t_3$
and $t_4$ Eq.~\eqref{eq:PotenialMoonIteration} with the
conditions $\phi_C(t_0)=$ 96~MV and 
$\Delta t = 10^{-6}$~s.
From $t_4$ on the Moon would stay at 1789~V until it enters again
the geomagnetic tail of the Earth. The time dependence of 
the Moon potential would be
periodically repeated in 27.3 days.

\section {Conclusion}

The absence of significant
macroscopic electrical fields in the universe 
supports the hypothesis of an electric neutral
(uncharged) universe. This
requests that all objects in the universe and the interactions
between them are electrical neutral. The intention
of our investigation was to test this hypothesis
locally in our solar system. The interaction 
between these neutral objects in our solar 
system is mainly caused by the
solar wind and cosmic rays. 
These objects will only stay uncharged if the 
impinging fluxes are a neutral plasma.

As test object we choose the Moon because the Moon has no
atmosphere, magnetic field, and extensive measurements
of the flux of solar wind, and cosmic rays are available in the
environment of the Moon.
Data for positive and negative charged particle fluxes exist
above 856 eV. In this
energy range used for our investigation the positive
charged flux is larger than the negative charged flux.
The integrated charge flux of the energy spectrum of the solar wind and
cosmic rays is accumulating to an excess of a total positive charge.
This causes a positive electric
potential of the Moon which is accompanied by an electric field.
The positive electric potential of the Moon
is depending on the position of the Moon on its orbit, between
1789~V outside the geomagnetic tail of the Earth and in maximum
96~MV inside this tail. Outside the geomagnetic tail where
the solar wind dominates,
the time $\Delta t $ needed to charge
or discharge the Moon is between 
$\Delta t = 13.4 ~\mu$s and $\Delta t = 1 $ms. 
Inside the
geomagnetic tail of the Earth the Moon will be charged up
in $\Delta t = 3.1 $ days to 96 MV.
If we consider an object in deep space, an object of the
size of the Moon would carry a positive electric potential
of the order of 4~GV, because the impact of the solar wind
will be negligible and the necessary time to charge the object would be
available. It is very likely that the cosmic rays exist not only
in our solar system, they exist with high probability in the
galactic system and the universe. As a consequence, every object
in the universe will carry an electric potential
originated by the cosmic rays.
These results violate substantial the hypothesis 
of an electric neutral universe.

The introduction of a so far unknown negative charged flux
$ Q(t)_{unid} $ in Eq.~\eqref{eq:PotentialMoon}
opens the possibility to set the potential of the Moon to zero.
Our investigations of the potential of the Moon 
put stringent conditions on $ Q(t)_{unid} $.
An excess of a negative charged particle flux
must exist in an energy range of $ 0 < E_{kin} < 856 $~eV.
This excess should be sensitive to the distance
between object and Sun to compensate for the decrease
of the flux of the solar wind with increasing distance
from Sun to object. A mechanism is required to compensate
for the short term changes of the of solar wind (flux),
for example, to eliminate the short term fluctuations
in the positive-negative charged flux of a solar flare.
A time-space depending modulation of this excess
would be able to stabilize the electric potential
of all objects in the universe to zero.
Finally, the excess should be insensitive to
the bow shock, heliopause and termination shock.

The energy of the universe is composed of
70$\%$ of dark energy,
26$\%$ of dark matter and 4$\%$ baryonic matter~\cite{vacuum}.
The baryonic matter in the universe is immersed in this electric
neutral environment. 
Currently extensive theoretical investigations
are performed to describe the dark energy and dark matter
as a condensate. For example as a Bose-Einstein condensate
\cite{BoseEinstein}, as flavor condensates \cite{flavorCondensates},
as a quantum Yang-Mills condensate \cite{Yang-Millscondensate} or
condensation of scalar fields \cite{condensationScalarField}.
If an electromagnetic interaction to such a condensate would
exist, in particular its superfluid characteristic
would allow to fulfill the above discussed 
difficult extensive conditions
to earth the Moon and all objects in the cosmos.
The energy content of an
excess of a negative charged particle flux
compared to the dark energy and energy of dark matter
is negligible. A perturbation of the condensate would be
minute.

If an object in space,
like the Earth, is surrounded by an atmospheric layer, it should
be possible to detect luminous electric discharges from the
Earth to the condensate (vacuum). Luminous electric discharges
like Flares or Transient Luminous Events \cite{Flares1,Flares2}
are observed in the environment of the Earth. Theories for
the majority of these discharges are based on the
local characteristic of these events in the Earth atmosphere.
For instance a theory of Elves in Earth-ionosphere
\cite{ELVES} or theory of Sprites \cite{SPITES1,SPITES2}.
So far exist no satisfactory theory of all experimental
observed luminous events. The
introduction of an electromagnetic interaction to the vacuum
could improve the theoretical understanding of this phenomena
and contribute to the discussion of the nature of dark energy
and dark matter.

\section*{Acknowledgments}
\noindent
We like to acknowledge the important support for 
this project of A. Badertscher and
A. Rubbia from the ETHZ Institute for Particle Physics
and the ETHZ Laboratory of Ion Beam Physics.
We thank the ACE instrument teams, the
ACE Science Center, the IMP-8, SOHO, and WIND instrument 
teams, and the CDAweb of the Goddard Flight Center for providing data.


\bibliographystyle{/l3/paper/biblio/l3stylem}
\bibliography{Vers23July10.bbl}

\end{document}